\documentclass[pra,a4paper]{revtex4}

\usepackage{amsmath}
\usepackage{graphicx}
\usepackage{epstopdf}

\newcommand{\bra}[1]{\mbox{$\langle #1|$}}
\newcommand{\ket}[1]{\mbox{$|#1\rangle$}}
\newcommand{\braket}[2]{\mbox{$\langle #1|#2\rangle$}}
\DeclareMathOperator{\Tr}{Tr}
  
\begin{document}

\title{Interface between path and OAM entanglement for high-dimensional photonic quantum information}

\author{Robert Fickler$^{1,2,\ast}$}
\author{Radek Lapkiewicz$^{1,2}$}
\author{Marcus Huber$^{3,4}$}
\author{Martin Lavery$^{5}$}
\author{Miles Padgett$^{5}$}
\author{Anton Zeilinger$^{1,2,\ast}$}

\affiliation{$^1$Vienna Center for Quantum Science and Technology, Faculty of Physics, University of Vienna,
Boltzmanngasse 5, Vienna A-1090, Austria
\\
$^2$Institute for Quantum Optics and Quantum Information, Austrian Academy of Sciences,
Boltzmanngasse 3, Vienna A-1090, Austria
\\
$^3$Fisica Teorica: Informacio i Fenomens Quantics, Departament de Fisica, Universitat Autonoma de Barcelona, E-08193 Bellaterra (Barcelona), Spain 
\\
$^4$ICFO-Institut de Ciencies Fotoniques, E-08860 Castelldefels (Barcelona), Spain
\\
$^5$School of Physics and Astronomy, Scottish Universities Physics Alliance (SUPA), University of Glasgow, Glasgow G12 8QQ, U.K.
\\
$^\ast$Correspondence to: robert.fickler@univie.ac.at and  anton.zeilinger@univie.ac.at}


\begin{abstract}
Photonics has become a mature field of quantum information science$^1$, where integrated optical circuits offer a way to scale the complexity of the setup as well as the dimensionality of the quantum state. On photonic chips, paths are the natural way to encode information$^2$. To distribute those high-dimensional quantum states over large distances, transverse spatial modes, like orbital angular momentum (OAM) possessing Laguerre Gauss modes$^3$, are favourable as flying information carriers$^{4,5,6,7}$. Here we demonstrate a quantum interface between these two vibrant photonic fields. We create three-dimensional path entanglement between two photons in a non-linear crystal and use a mode sorter$^{8,9}$ as the quantum interface to transfer the entanglement to the OAM degree of freedom. Thus our results show a novel, flexible way to create high-dimensional spatial mode entanglement. Moreover, they pave the way to implement broad complex quantum networks$^{10}$ where high-dimensionally entangled states could be distributed over distant photonic chips.
\end{abstract}

\date{\today}

\maketitle

Integrating optical elements onto photonic chips enables the scaling of complex experiments. The, path of the photons, a property that is inherently extendable to higher-dimensional systems$^{2,11}$, is widely used to encode information. Another possibility of encoding high-dimensional quantum information is the transverse spatial mode of light$^3$. Various modes, like OAM carrying Laguerre-Gauss (LG)$^{12,13}$, Ince-Gauss$^{14}$ and Bessel-Gauss$^{15}$ have been used to demonstrate experimentally high-dimensionally entangled states and to implement quantum informational tasks. Recently, first experiments of an integrated OAM beam emitter interconnected the free-space and chip-based fields although not yet in the quantum regime$^{16}$.
 
In this letter we demonstrate a quantum interface between these two approaches to high-dimensional photonic quantum information: path encoding for complex on-chip experiments and OAM carrying light modes to transmit the information over large distances. At the same time, we investigate a novel, flexible way to create higher-dimensional OAM entanglement not relying on angular momentum correlation. The essential tool in our experiment is a mode sorter (MS) which was invented$^{8,9}$ to convert the OAM content of an incident light beam to lateral positions of the output beam. By using this device in reverse, different spatial positions are transformed to the cylindrically symmetric LG modes. We demonstrate that a high-dimensional path-entangled state can be transferred to high-dimensional OAM entanglement with the help of the reversed MS.
\begin{figure}
\centering  \includegraphics[width=0.6\textwidth ]{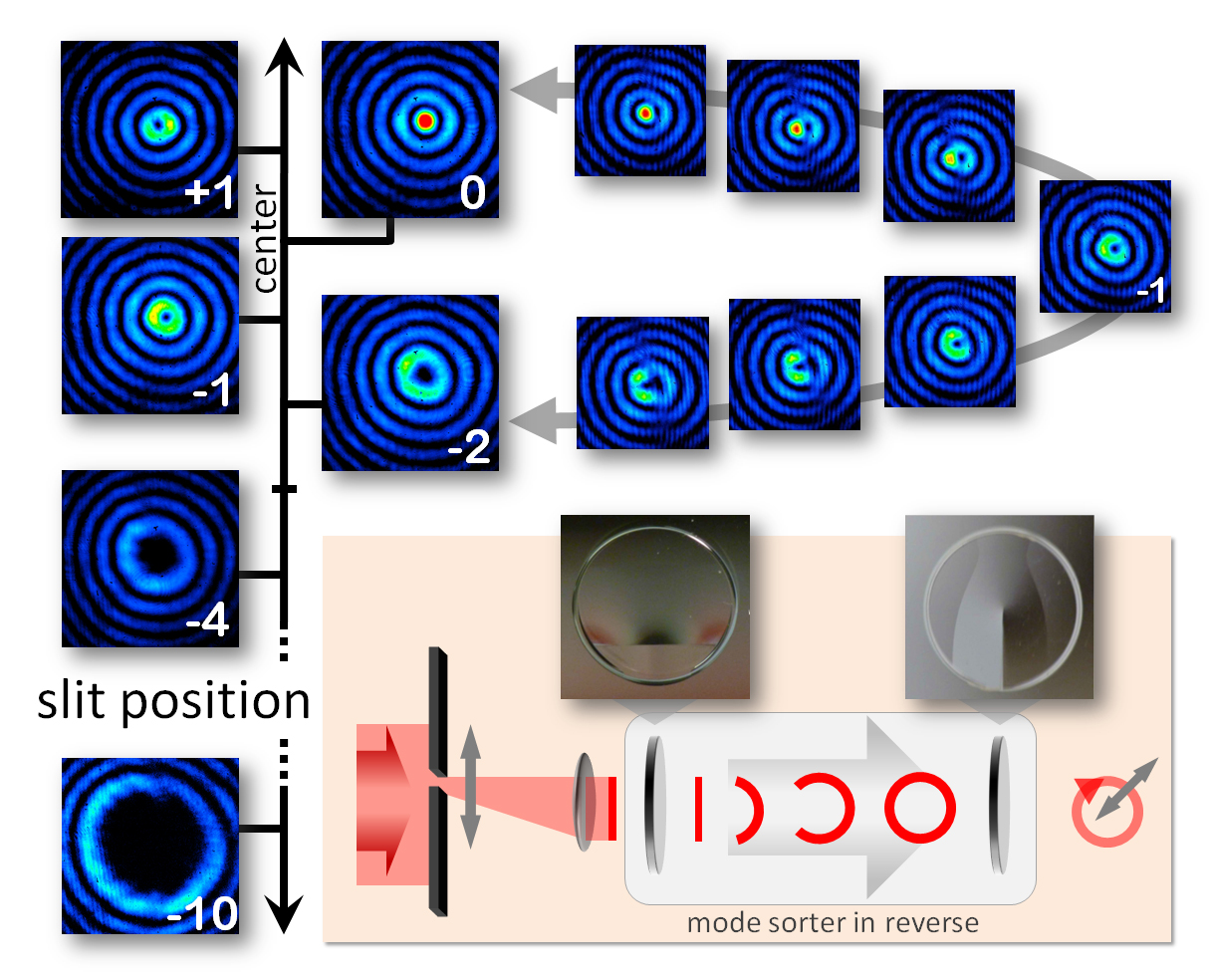}
\caption{Generation of OAM states from lateral position states. False colour images recorded with a CCD camera are shown for various OAM states up to the 10$^{th}$ order (white numbers). The distance between the lateral positions of the slit that generate integer OAM states was around 150 $\mu$m for the shown images. In between these positions, fractional OAM modes are generated. The multiple ring structure arises, because the light diffracted of the slit is transformed by the MS to higher-order radial modes containing the same OAM. Inset: Sketch of the setup, where light is diffracted through a movable slit (width approx. 100µm) positioned in the focal plane of a lens. It is transformed by two freeform refractive optical elements (MS - mode sorter in reverse) into different OAM states depending on the slit position (detailed description in the main text). \label{fig1}}
\end{figure}

Transverse spatial LG modes have a helical phase front $e^{il\theta}$, where $l$ can take any integer value and represents the quanta of OAM each individual photon possesses$^{17}$. If $l\neq 0$, such modes exhibit a vortex along the beam axis and show a ring-shaped intensity structure, consequently they are also called ``doughnut modes". LG modes can be used to transmit more classical information$^{4,5,6,7}$ or realise a higher-dimensional quantum state$^{13,18,19}$. To access the encoded information efficiently a MS was developed$^{8,9}$ which consists of two freeform refractive optical elements that convert the OAM content of an incident light field to lateral positions of the output. The first element maps the azimuthal to the lateral coordinates. Thereby, the ring-shaped intensity is transferred to a straight intensity line and the $l$-dependent helical phase structure to a transverse phase gradient. The second element corrects for phase distortion due to optical path length differences. A lens after the second element Fourier-transforms the transverse phase gradient to specific spatial position i.e. finishes the sorting of the modes. Recently, the operation of the MS in reverse was demonstrated by converting a light field with a transverse phase gradient, which has been created with a spatial light modulator (SLM), into an LG mode$^{20}$.

\begin{figure}
\centering  \includegraphics[width=0.6\textwidth ]{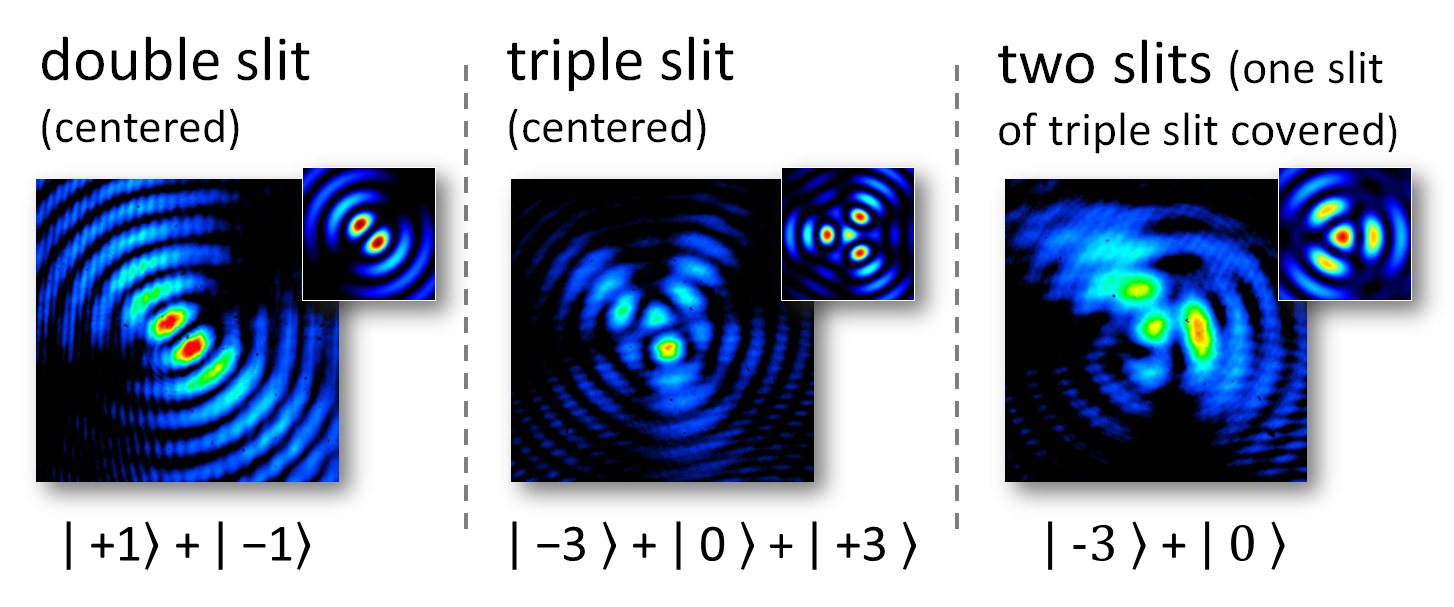}
\caption{Creation of superpositions of OAM states. If a double or triple slit is placed in front of the MS (Fig \ref{fig1} inset) different superpositions of OAM modes are created (false colour images) dependent on the slit dimensions and focusing parameters of the lens (see methods). All imaged modes show reasonable qualitative agreement with the theoretical calculation of the expected superpositions (insets). \label{fig2}}
\end{figure}

Here, we investigate a different, simple way of using the mode sorter operated in reverse as a source for OAM states. A narrow slit that diffracts the light is positioned in the focal plane of the lens (Fig. \ref{fig1} inset). Thus, behind the lens, a parallel beam emerges with a phase profile given by the position of the slit relative to the optical axis. That phase distribution can be adjusted to be a multiple of $2\pi$ by adjusting the lateral position of the slit. A subsequent MS in reverse transforms the state's amplitude distribution into a circle and thus into an OAM state (Fig. \ref{fig1} inset). A continuous lateral movement of the slit leads to integer OAM modes at multiples of $2\pi$ and fractional OAM in intermediate positions (see Fig. \ref{fig1}). We were able to create LG modes up to order $l=\pm 10$. The order of the generated modes was only limited by the size of the laser beam used to illuminate the slit. Importantly, by using multiple slits superpositions of LG modes could also be generated$^{21}$ (Fig. \ref{fig2}). A qualitative confirmation that the generated annular shaped modes have the correct helical phase, thus the expected OAM value, can be found in supplementary section (Fig. \ref{figS1}).

\begin{figure}
\centering  \includegraphics[width=0.80\textwidth ]{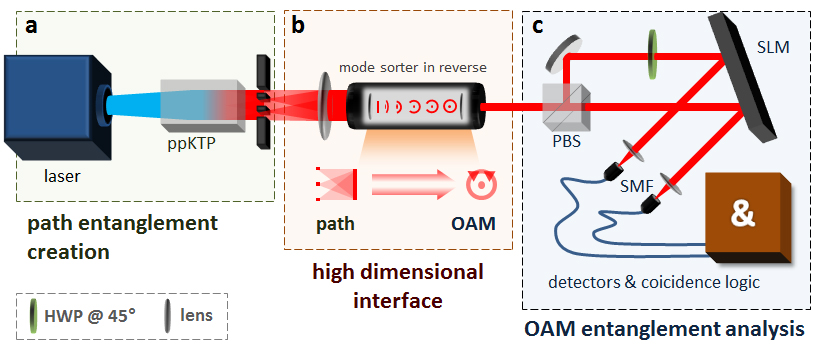}
\caption{Schematic sketch of the experimental setup to interface high-dimensional path and OAM entanglement. a) By pumping a non-linear crystal (ppKTP) with a 405 nm laser (blue) we create pairs of orthogonally polarized photons with 810 nm wavelength in a down conversion process (red). A triple slit is placed directly behind the crystal. Both photons of a pair emerge together through the one of the three slits. The final state is a superposition of the pair passing the three slits hence the pair is entangled in path. b) A lens transforms the path entanglement to a transverse momentum entanglement which is then transformed by the reversed mode sorter (black tube) to the OAM degree of freedom. c) The OAM-entangled pair of photons is split by a polarizing beam splitter (PBS) and the vertically polarized photon is rotated by a half-wave plate (HWP) to the required polarization for phase modulation with the SLM. Each photon is filtered individually depending on its spatial mode, by a hologram displayed on a spatial light modulator (SLM) and by coupling to single-mode fibres (SMF). Avalanche photo-diode, single-photon detectors and a coincidence-logic (brown box) are utilized to detect photon pairs and the observed correlations are subsequently analysed. \label{fig3}}
\end{figure}

In this present work, we unambiguously demonstrate that the described method can be used to create single photons carrying OAM and, more importantly, to transform a bipartite path-entangled state to an OAM-entangled state, even for higher-dimensional states. We create pairs of orthogonally polarized, path-entangled photons from position correlation in a spontaneous parametric down-conversion process$^{11,22,23}$ (Fig. \ref{fig3} a and methods). By placing a triple slit directly behind the crystal we filter out only three positions which leads to an expected two-photon qutrit path-entangled state
\begin{eqnarray}
\ket{\psi}=a\ket{S_1}_H\ket{S_1}_V+e^{i\phi_1}b\ket{S_2}_H\ket{S_2}_V+e^{i\phi_2}c\ket{S_3}_H\ket{S_3}_V \quad ,\label{eq:State}
\end{eqnarray}
where $S_{1/2/3}$ denote the slit that both photons pass and the subscripts $H/V$ label their polarization. The amplitudes $a$, $b$ and $c$ ($a^2+b^2+c^2=1$) as well as the phases $\phi_1$ and $\phi_2$ are described by real numbers and depend on the pump beam behind each slit. Therefore, they are flexibly adjustable by modulating the intensity and phase of the pump beam. If only two slits are used a state is created which consists of the first two terms.

The slits are placed in the focal plane of a lens followed by the reversed MS which leads to a transformation of the path entanglement to the OAM degree of freedom. Thus, the total transformation acts as an interface between high-dimensional path and spatial mode entanglement (Fig \ref{fig3} b). In our experiment the distance between the three (two) slits correspond to the 0$^{th}$, -3$^{rd}$  and +3$^{rd}$ order LG modes (0$^{th}$ and -3$^{rd}$ order). The third order modes are chosen to reduce the modal overlap due the MS to less than 1\%. Hence the expected state can be written as 
\begin{eqnarray}
\ket{\psi}=a\ket{-3}_H\ket{-3}_V+e^{i\phi_1}b\ket{0}_H\ket{0}_V+e^{i\phi_2}c\ket{+3}_H\ket{+3}_V \quad ,\label{eq:state_final}
\end{eqnarray}
where $-3,0,3$ label the order of the mode and OAM quanta respectively. The flexibility in adjusting the amplitudes, phases, OAM values and dimensionality of the state (via transmittance, positions and number of slits) implies a general method to custom-tailor high-dimensional OAM entanglement. To verify the OAM entanglement we split the transferred photon pair with a polarizing beam splitter and analyse each photon with a spatial mode filter. The filter is realised by a combination of a phase hologram on the SLM and a single-mode fibre which only couples fundamental Gauss modes$^{12,24}$. Single-photon detectors (avalanche diodes) together with a coincidence-logic are used to register correlations between the two spatial modes of a pair (Fig \ref{fig3} c). 

\begin{figure}
\centering  \includegraphics[width=0.6\textwidth ]{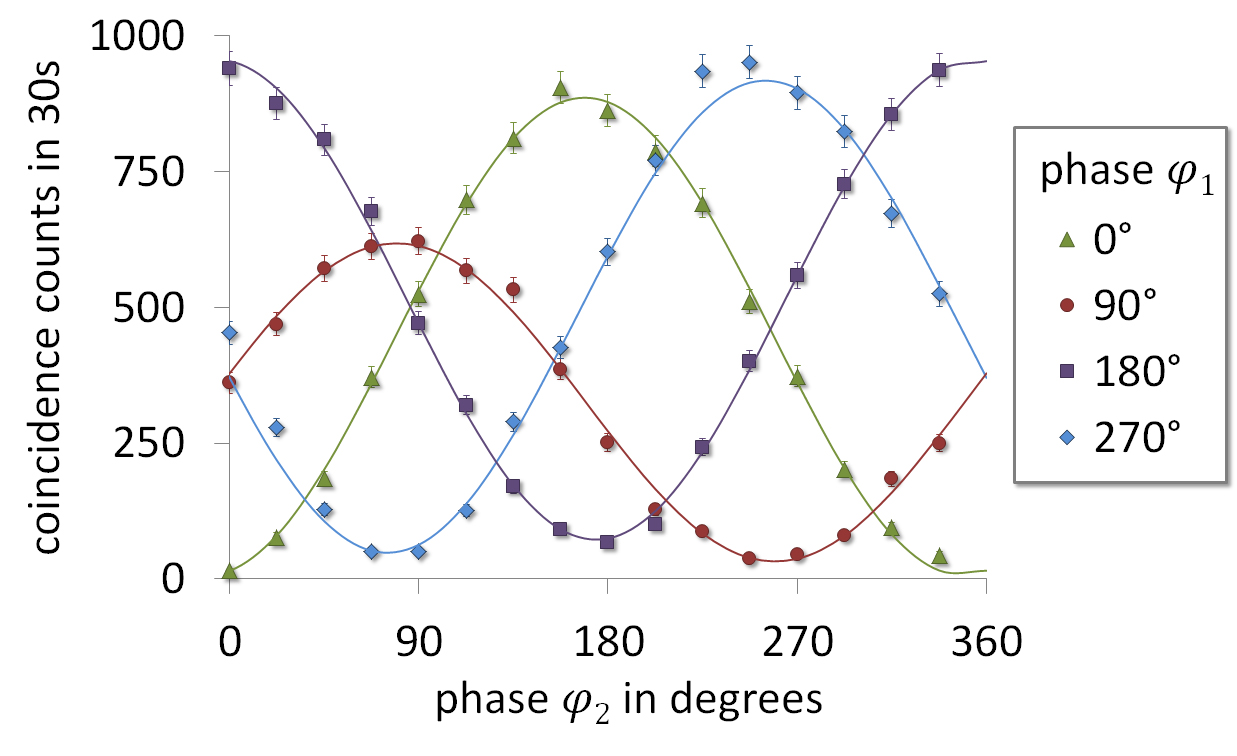}
\caption{Coincidence measurements to confirm the two-dimensional OAM entanglement. With the hologram displayed on the SLM we measured for each observed photon the superposition $\ket{-3}+e^{i\varphi}\ket{0}$, where $\varphi$ is an adjustable phase between $0^\circ$ and $360^\circ$. We measured coincidence fringes for 4 different settings of the phase $\varphi_1$ of one photon while the phase $\varphi_2$ setting for the partner photon is scanned in steps of $22.5^\circ$. The lines show the best $\sin^2$-fits which lead to visibilities of around $90\%$, confirming entanglement. For one setting ($90^\circ$) the maximum is found to be significantly smaller then for the other phase settings. This might originate from a slight misalignment of the mode sorter. Error bars (if big enough to be seen) depict one sigma confidence with Poissionian count statistics assumed.  \label{fig4}}
\end{figure}

We quantitatively demonstrate entanglement by using a simple, powerful entanglement witness that compares the extracted fidelity (overlap between an ideal state of equation (\ref{eq:state_final}) with the measured data) and the maximally expected fidelity for d-dimensional entangled states. If the measured fidelity exceeds the known bound for a d-dimensional entangled state, our results prove at least (d+1)-dimensional entanglement (see supplementary section). In a first experiment we test for qubit-entanglement by measuring correlations between the 0$^{th}$ order (Gauss mode) and the -3$^{rd}$ order LG mode (Fig. \ref{fig4}). The highest fidelity calculated from the measured visibilities in all mutually unbiased bases was $97\pm 2\%$ with an ideal state where $a=0.54$, $b=0.84$. The maximum fidelity that would be achievable for separable states is $71\%$. This confirms our observation of two-dimensional entanglement. Additionally, a Bell-CHSH-type test for qubit-entanglement$^{25}$ was successfully performed with a statistical significance of more than 10 standard deviations (for more details see supplementary section). In a second experiment we take advantage of all three implemented slits and test for OAM qutrit-entanglement. Similar to earlier results$^{14,19}$, the entanglement witness enables us to draw conclusions about the global dimensionality of the entanglement whilst restricting ourselves to qubit-subspace measurements. In our experiment this approach corresponds to the measurements of all visibilities in all two-dimensional subspaces that is for l-values of $0 / -3$, $0 / +3$ and $-3 / +3$. The best statistical significance for genuine qutrit-entanglement was found for an ideal state where $a=c=0.48$, $b=0.73$. Here, the fidelity obtained from the measured visibilities is $89\pm 4\%$, which exceeds the upper bound for any two-dimensional entangled state ($77\%$) by more than three standard deviations. Note that in both cases (qubit and qutrit-entanglement) no background subtraction has been applied. Also, the largest amplitude was found for the Gauss mode because it corresponds to the central slit where the pump intensity and thus the down conversion rate is maximal.

Our setup is not limited to qutrits but can be naturally extended to d-dimensional entanglement. A broader pump beam and a wider crystal in combination with a d-slits arrangement would lead to d possible paths and thus d-dimensional entanglement. A further way to increase the dimensionality of the path entanglement as well as the efficiency would be an arrangement with many integrated down-conversion crystals$^{26}$ in waveguide structures or fibre-coupled down-conversion crystals pumped in parallel$^{2}$. In that case the waveguides or a fibre groove array could replace the multiple slits. Furthermore it has been shown recently that the MS works for up to 50 states$^{27}$ and that the MS can be improved to reduce the overlap between neighbouring modes$^{28}$.  These improvements in MS design suggest that our approach can be readily extended to higher qudit entanglement. 
In conclusion, we have shown that the mode sorter in combination with a slit can be used, in reverse, to generate LG modes and their superpositions up to at least $l =\pm10$. Using fast switching techniques for optical paths the method can increase the switching rate between different LG beams in classical information technologies or LG mode based quantum cryptography applications. In addition, our results demonstrate a novel, flexible way to create high-dimensional OAM entanglement. Most importantly, they show a way to implement an quantum interface between two approaches to high-dimensional quantum information: path encoding in waveguide structures for scalable, complex photonic quantum circuits including arbitrary unitary operation for higher-dimensional states$^{1,2}$ and OAM encoding to distribute those high-dimensional quantum states$^{4,5,7}$ in a broad quantum network scenario.

\section{Methods}
\textbf{Creation of path-entangled photons}\\
To create correlated photons we used a cw 405 nm laser diode to pump a periodically poled potassium titanyl phosphate (ppKTP) crystal (1x2x5 mm) with approx. 30 mW pump power. The pump beam diameter was approx. 1 mm (FWHM) which leads to a broad region where the two orthogonally polarized photons (Type II down conversion) with 810 nm wavelength can be created. In the near-field of the crystal – directly behind it –  the photons are correlated in the transverse spatial position$^{21,22}$, i.e. if one photon is found to be at a specific transverse spatial location the partner photon will be found ideally at exactly the same location too. This stems from the down conversion process itself where the pump photon generates two down converted photons at some transverse spatial position of the crystal. Because the actual location of the down conversion process is not determined, i.e. it can happen everywhere within the pump beam spread, the generated photon pairs are in a superposition of all possible positions. A double or triple slit placed behind the crystal selects only two or three lateral positions, thus the photon pairs that pass the slits are in a superposition of two or three locations. In other words, they are in path-entangled qubit or qutrit states. The slits had a width of 150µm and were separated by 250 $\mu$m which resemble the dimensions typically used for telecom-standard fibre arrays. A schematic sketch of the setup can be seen in figure \ref{fig3} a. The presented scheme is readily extendable to more than three-dimensional entanglement, by increasing the number of slits. Furthermore our source could be directly connected to complex integrated waveguide structures. 

\section{Acknowledgements}
The authors thank Sven Ramelow and Mario Krenn for fruitful discussions, Milan Mosonyi for helping with the proof of the witness and Otfried G\"uhne for giving valuable insight into the solution, by pointing out that the overlap in eq. (\ref{eq:guehne}) was already proved (in a different way) in his PhD-Thesis. 
This work was supported by the Austrian Science Fund (FWF) through the Special Research Program (SFB) Foundations and Applications of Quantum Science (FoQuS; Project No. F4006-N16) and the European Community Framework Programme 7 (SIQS, collaborative project, 600645). RF and RL are supported by the Vienna Doctoral Program on Complex Quantum Systems (CoQuS, W1210-2). MH would like to acknowledge the MarieCurie IEF grant QuaCoCoS – 302021. MPJL and MJP are supported by the EPSRC.

\section{References}
1.	O'Brien, J.L., Furusawa A. \& Vuckovic J. Photonic quantum technologies. \textit{Nature Photonics} \textbf{3}, 687-695 (2009).\\
2.	Schaeff, C. et al. Scalable fiber integrated source for higher-dimensional path-entangled photonic quNits. \textit{Optics Express} \textbf{20}, 16145-16153 (2012).\\
3.	Andrews, D. L. \& M. Babiker, eds. The angular momentum of light (Cambridge University Press, Cambridge, 2012).\\
4.	Gibson, G. et al. Free-space information transfer using light beams carrying orbital angular momentum. \textit{Optics Express} \textbf{12}, 5448-5456 (2004).\\
5.	Tamburini, F. et al. Encoding many channels on the same frequency through radio vorticity: first experimental test. \textit{New Journal of Physics} \textbf{14}, 033001 (2012).\\
6.	Wang, J. et al. Terabit free-space data transmission employing orbital angular momentum multiplexing. \textit{Nature Photonics} \textbf{6}, 488-496 (2012).\\
7.	Bozinovic, N. et al. Terabit-Scale Orbital Angular Momentum Mode Division Multiplexing in Fibers. Science 340, 1545-1548 (2013).\\
8.	Berkhout, G.C., Lavery, M.P., Courtial, J., Beijersbergen, M.W. \& Padgett, M.J. Efficient sorting of orbital angular momentum states of light. \textit{Physical Review Letters} \textbf{105}, 153601 (2010).\\
9.	Lavery, M.P.J. et al. Refractive elements for the measurement of the orbital angular momentum of a single photon. \textit{Optics Express} \textbf{20}, 2110-2115 (2012).\\
10.	Kimble, H.J. The quantum internet. \textit{Nature} \textbf{453}, 1023-1030 (2008).\\
11.	Rossi, A., Vallone, G., Chiuri, A., De Martini, F. \& Mataloni, P. Multipath entanglement of two photons. \textit{Physical Review Letters} \textbf{102}, 153902 (2009).\\
12.	Mair, A., Vaziri, A., Weihs, G. \& Zeilinger, A. Entanglement of the orbital angular momentum states of photons. \textit{Nature} \textbf{412}, 313-316 (2001).\\
13.	Vaziri, A., Weihs, G. \& Zeilinger A. Experimental two-photon, three-dimensional entanglement for quantum communication. \textit{Physical Review Letters} \textbf{89}, 240401 (2002).\\
14.	Krenn, M. et al. Entangled singularity patterns of photons in Ince-Gauss modes. \textit{Physical Review A} \textbf{87}, 012326 (2013).\\
15.	McLaren, M. et al. Entangled Bessel-Gaussian beams. \textit{Optics Express} \textbf{20}, 23589-23597 (2012).\\
16.	Cai, X. et al. Integrated compact optical vortex beam emitters. \textit{Science} \textbf{338}, 363-366 (2012).\\
17.	Allen, L. et al. Orbital angular momentum of light and the transformation of Laguerre-Gaussian laser modes. \textit{Physical Review A} \textbf{45}, 8185-8189 (1992).\\
18.	Dada, A.C., Leach, J., Buller, G.S., Padgett, M.J. \& Andersson, E. Experimental high-dimensional two-photon entanglement and violations of generalized Bell inequalities. \textit{Nature Physics} \textbf{7}, 677-680 (2011).\\
19.	Krenn, M. et al. Studies of Quantum Entanglement in 100 Dimensions. arXiv preprint arXiv:1306.0096 (2013).\\
20.	Lavery, M.P.J. et al. The measurement and generation of orbital angular momentum using an optical geometric transformation. SPIE LASE. International Society for Optics and Photonics, (2013).\\
21.	Berkhout, G.C., Lavery, M.P., Padgett, M.J. \& Beijersbergen, M.W. Measuring orbital angular momentum superpositions of light by mode transformation. \textit{Optics Letters} \textbf{36}, 1863-1865 (2011).\\
22.	Edgar, M.P. et al. Imaging high-dimensional spatial entanglement with a camera. \textit{Nature Communications} \textbf{3}, 984 (2012).\\
23.	Moreau, P.A., Mougin-Sisini, J., Devaux, F. \& Lantz, E. Realization of the purely spatial Einstein-Podolsky-Rosen paradox in full-field images of spontaneous parametric down-conversion. \textit{Physical Review A} \textbf{86}, 010101 (2012).\\
24.	Leach, J. et al. Violation of a Bell inequality in two-dimensional orbital angular momentum state-spaces. \textit{Optics Express} \textbf{17}, 8287-8293 (2009).\\
25.	Clauser, J.F., Horne, M.A., Shimony, A. \& Holt, R.A. Proposed experiment to test local hidden-variable theories. \textit{Physical Review Letters} \textbf{23}, 880-884 (1969).\\
26.	Harder, G. et al. An optimized photon pair source for quantum circuits. \textit{Optics Express} \textbf{21}, 13975-13985 (2013).\\
27.	Lavery, M.P.J. et al. Efficient measurement of an optical orbital-angular-momentum spectrum comprising more than 50 states. \textit{New Journal of Physics} \textbf{15}, 013024 (2013).\\
28.	Mirhosseini, M., Malik, M., Shi, Z. \& Boyd, R.W. Efficient separation of the orbital angular momentum eigenstates of light. \textit{Nature Communications} \textbf{4} (2013). \\

\newpage

\section{Supplementary}

\subsection{Qualitative test of the phase front of the generated OAM beam}
In the main text we describe the unambiguous demonstration that annular shaped modes generated by the mode sorter (MS) in reverse carry the expected OAM value. We do this by spatial mode filtering, implemented by holographic transformation followed by a single mode fibre that only collects the fundamental Gauss mode, in the measurement to test for entanglement. Nevertheless, it is interesting to test more qualitatively the generated phase of the beam too. For this, we generate an OAM laser beam as described in figure 1 of the main text. Afterwards, we modulate the modes with the opposite handed phase holograms displayed on a SLM and imaged the beam with a CCD camera (Fig. S1). The recorded modal structure shows an intensity along the beam axis in all three tests which resembles a fundamental Gauss mode (or 0$^{th}$ order Bessel beam containing no OAM). However, a reduction of the Gauss mode quality can be seen for higher orders, which is in agreement with the count rates of the entanglement experiment presented in the main text (lower count rates have been measured for $l=\pm3$).
\begin{figure}[bht]
\centering  \includegraphics[width=0.48\textwidth ]{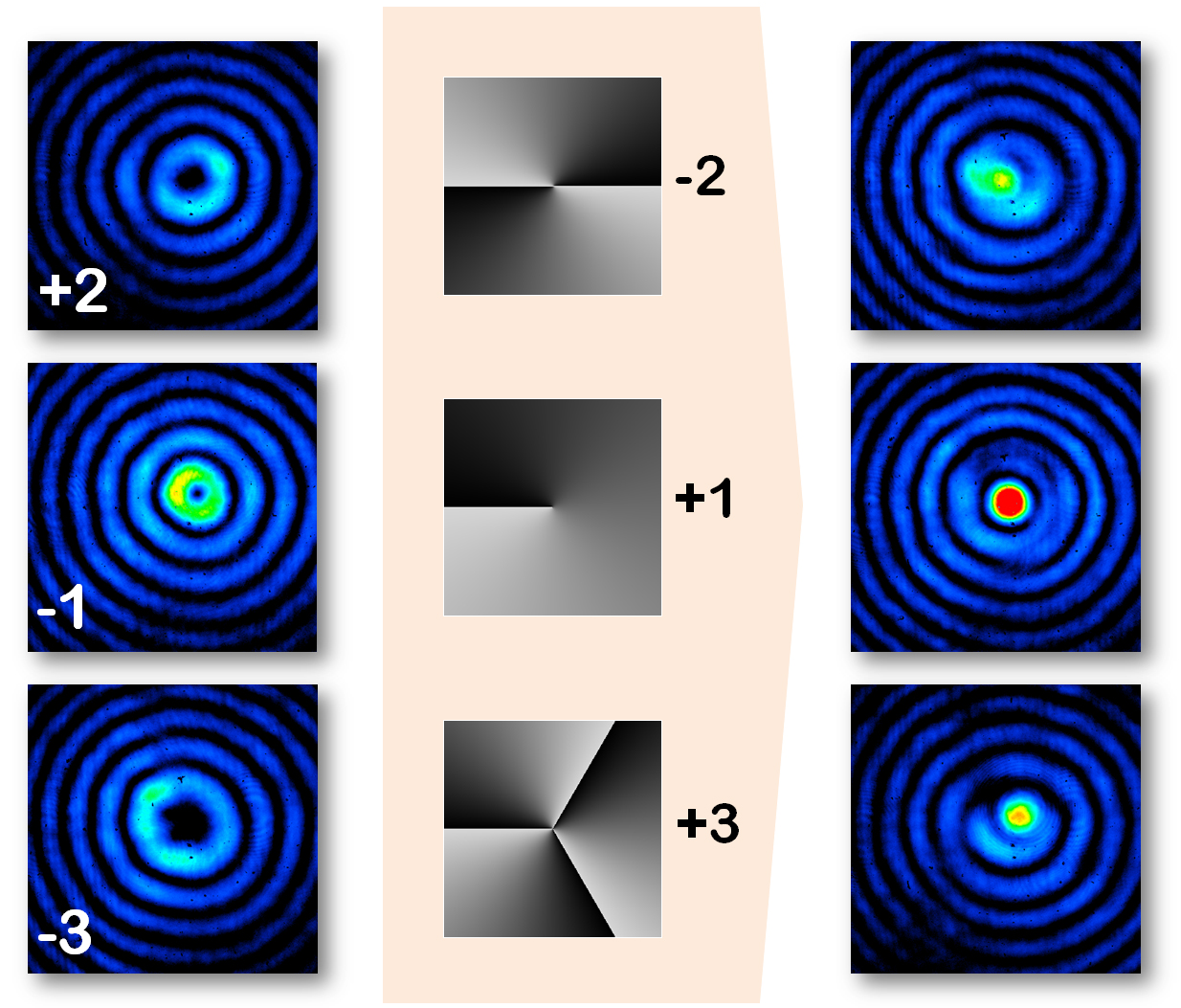}
\caption{Examination of phase structure of the generated mode. We use an SLM to modulate the generated mode (left side) with the opposite handed phase hologram (phase modulation shown in grey scale, linearly from 0=black to 2$\pi$=white). The resulting intensities (right side) resemble the expected fundamental Gauss mode with an on axis intensity maximum, thus confirming the successful generation of helically phased OAM modes. We verified these results in the entanglement experiment described in the main text by coupling this mode into a single-mode fibre. \label{figS1}}
\end{figure}  

\subsection{Bell-CHSH test for qubit entanglement}
In addition to the results of the entanglement witness (theoretical proof in the next chapter) presented in the main text, we used the measurements shown in figure 4 of the main text to test for two-dimensional entanglement with the popular criterion of a CHSH-Bell-inequality (reference 26 in the main text). The generated qubit-entangled state was assumed to be $\ket{-3,-3}+\ket{0,0}$, where the numbers label the OAM value. For local realistic theories the following bound holds
\begin{equation}
S_{CHSH} = \left| E(\alpha ,\beta ) - E(\alpha' ,\beta ) + E(\alpha ,\beta' ) + E(\alpha' ,\beta' ) \right| \leq 2 , 
\end{equation}
where $\alpha$, $\alpha'$, $\beta$ and $\beta'$ denote different measurement settings (phases of the measured superpositions) and E stands for the normalized expectation value for photon pairs to be found with this combinations of modes. In our measurement we achieve a value of 2.47 $\pm$ 0.04 which violates the classical bound by more than ten standard deviations (Poissionian count statistics assumed), thus verifying qubit-entanglement.

\subsection{Proof of the bipartite witness for d-dimensional entanglement}

We develop a witness framework requiring only measurements in two-dimensional subspaces. From these subspace measurements we compute the fidelity $F$ of the experimentally produced state with a specially chosen high-dimensional state. We then use the techniques developed below to bound the maximal overlap of states with a bounded Schmidt rank $d$ and the chosen high dimensional state. If the subspace measurements reveal a higher overlap than this bound, the production of at least $d+1$-dimensional entanglement is proven.

\noindent \textbf{Construction of the witness}\\
The Schmidt decomposition of the assumed high dimensional state can be written as $\ket{\psi}=\sum_{i=0}^{D-1} \lambda_{i} \ket{ii}$, where $i$ labels the different states (in our experiment the spatial modes), $D$ denotes the dimension of the Hilbert space and the coefficients are chosen in a decreasing order, i.e. $\lambda_1 \geq \lambda_2 \geq (...) \geq \lambda_d$ (in the main text ${\lambda_i}$ corresponds to the amplitudes $a$, $b$ and $c$). This choice has the advantage that $\text{Tr}(\rho\ket{\psi}\bra{\psi})=\sum_{i,j=0}^{D-1}\frac{1}{\lambda_i\lambda_j}\langle ii|\rho|jj\rangle$, where all appearing matrix elements can be determined by three visibility measurements in two-dimensional subspaces, the number of which scaling as the square root of those required for a full state tomography. We can now construct a witness for $d$-dimensional entanglement (Schmidt number witness) by comparing the two fidelities
\begin{eqnarray}
F &=& \text{Tr}(\rho \ket{\psi}\bra{\psi}) \label{eq:witness}\\
f_d&=& \max_{\ket{\phi_{d}}} |\braket{\phi_{d}}{\psi}|^2 \label{eq:witness2}
\end{eqnarray}
where $\rho$ labels the density matrix related to our measurements and $\ket{\phi_{d}}$ denotes states with a bounded Schmidt rank $d$. If equation $F>f_d$ holds, the measurements can not be explained by a $d$-dimensional entangled state, thus the generated bipartite system was (at least) genuinely $(d+1)$-dimensionally entangled.

\noindent \textbf{Calculation of the maximal fidelity for d-entangled states}\\
The maximization in equation \ref{eq:witness2} runs over all states with at most $d$-dimensional entanglement (Schmidt rank $d$). This maximal overlap between $\ket{\phi_d}=\sum_{k,l=0}^{D-1} c_{kl} \ket{kl}$ and the guessed state $\ket{\psi}$ (expressed in terms of the Schmidt coefficients) can be rewritten to
\begin{eqnarray}
f_{d} 	&=& \max_{\ket{\phi_{d}}} \left| \left( \bra{kl}\sum_{k,l=0}^{D-1} c_{kl}^* \right) \left( \sum_{i=0}^{D-1} \lambda_{i} \ket{ii}\right) \right|^2 \\
		&=& \max_{\ket{\phi_{d}}} \left|\sum_{k,l,i=0}^{D-1}\braket{k}{i}\braket{l}{i}c_{kl}^*\lambda_i\right|^2 .
\end{eqnarray}
By rearranging the terms and introducing the operator $B=c_{kl}\ket{k}\bra{l}$ 
\begin{eqnarray}
f_{d} &=& \max_{\ket{\phi_{d}}} \left|\Tr \left( B^\dagger \sum_{i=0}^{D-1}\lambda_i \ket{i} \bra{i} \right)\right|^2
\end{eqnarray}
and the rank $d$-projector $P_d B^{\ast}=B^{\ast}$ (which always exists if $B{^\ast}$ is of rank $d$, as $\ket{\phi_{d}}$ is of Schmidt rank $d$) we get
\begin{eqnarray} 		
		&=& \max_{\ket{\phi_{d}}} \left|\Tr \left( P_d B^\dagger \sum_{i=0}^{D-1}\lambda_i \ket{i} \bra{i} \right)\right|^2 ,
\end{eqnarray}
and since the trace is invariant under cyclic permutations we can write
\begin{eqnarray}
f_{d} 	&=& \max_{\ket{\phi_{d}}} \left|\Tr \left( B^\dagger \sum_{i=0}^{D-1}\lambda_i \ket{i} \bra{i} P_d \right)\right|^2
\end{eqnarray}
Using the Cauchy-Schwarz inequality for the Hilbert-Schmidt inner product (for the inner product $\langle A,B\rangle:=\text{Tr}(AB^\dagger)$ it reads $|\langle A,B\rangle|^2\leq\langle A,A\rangle\langle B,B\rangle$) we get the inequality
\begin{eqnarray}
f_{d} 	&\leq & \max_{\ket{\phi_{d}}} \Tr (B^\dagger B) \Tr \left(P_d\sum_{i=0}^{D-1}{\lambda_i}^2 \ket{i} \bra{i} P_d \right) .
\end{eqnarray}
Because $\Tr B^{\dagger}B=\sum_{k,l=0}^{D-1}c_{kl}c_{lk}^*\leq 1$ and choosing the obviously maximizing $P=\sum_{i=0}^{d-1}\ket{i}\bra{i}$ we get the upper bound for the fidelity of $d$-dimensional entangled states of
\begin{eqnarray}
f_{d} 	&\leq & \sum_{i=0}^{d-1}{\lambda_i}^2 .
\end{eqnarray}
By choosing $\ket{\phi_{d}}=\ket{\Phi}=\frac{1}{\sqrt{\sum_{i=0}^{d-1}\lambda_i^2}}\sum_{i=0}^{d-1}\lambda_i|ii\rangle$ we find that $f_d\geq \sum_{i=0}^{d-1}{\lambda_i}^2$, thus proving that our bound is indeed tight, i.e.
\begin{align}
f_d= \max_{\ket{\phi_{d}}} |\braket{\phi_{d}}{\psi}|^2=\sum_{i=0}^{d-1}{\lambda_i}^2 . \label{eq:guehne}
\end{align}
\noindent \textbf{Inequality for the witness of d-dimensional entanglement}\\
Hence, from the visibility measurements in all two-dimensional subspaces we can now reveal information about the global dimensionality of the bipartite entanglement. If the following inequality holds, we have proven that our measurement results can only be explained by an at least $(d+1)$-dimensionally entangled state.
\begin{eqnarray}
\Tr (\rho \ket{\psi}\bra{\psi}) &>& \sum_{i=0}^{d-1}{\lambda_i}^2  \label{eq:witness3} 
\end{eqnarray}
Note that although we have derived the proof by assuming a pure state the witness holds even for mixed states because they would only lower the bound (due to the convexity of the fidelity). Thus, the presented witness is a state independent test for high-dimensional entanglement.


\end{document}